\documentclass[twocolumn, tighten, times]{aastex701}
\usepackage{amsmath}
\usepackage[caption=false]{subfig}
\usepackage{multirow}
\usepackage{booktabs}
\usepackage{array}
\usepackage{tabularx}
\usepackage{verbatim}
\usepackage{float}
\usepackage{soul}
\usepackage[normalem]{ulem}
\usepackage{lineno}

\newcommand{\rhog}{$n_{\rm \rm gd}$}

\newcommand{\D}{${\cal D}$}

\newcommand{\q}{$q$}

\newcommand{\sm}{$M_{\rm \odot}$}
\newcommand{\cm}{cm$^3$}
\newcommand{\dg}{$^{\circ}$}

\shorttitle{Black Hole Pairs on Inclined Orbits}
\shortauthors{Ghobadi et al.}

\begin{document}
\
\title{Evolution of Supermassive Black Hole Pairs on Inclined Orbits in Post-Merger Galaxies} 

\author[0009-0003-3513-1887]{Sena Ghobadi}
\affiliation{School of Physics and Center for Relativistic
  Astrophysics, 837 State St NW, Georgia Institute of Technology,
  Atlanta, GA 30332, USA}   
\email{sghobadi6@gatech.edu}

\author[0000-0001-8128-6976]{David R. Ballantyne}
\affiliation{School of Physics and Center for Relativistic
  Astrophysics, 837 State St NW, Georgia Institute of Technology,
  Atlanta, GA 30332, USA}
\email{david.ballantyne@physics.gatech.edu}

\author[0000-0002-7835-7814]{Tamara Bogdanovi{\'c}}
\affiliation{School of Physics and Center for Relativistic
  Astrophysics, 837 State St NW, Georgia Institute of Technology,
  Atlanta, GA 30332, USA}
\email{tamarab@gatech.edu}

\correspondingauthor{Sena Ghobadi}
\email{sghobadi6@gatech.edu}
\email{david.ballantyne@physics.gatech.edu}
\email{tamarab@gatech.edu}
\begin{abstract}
Theoretical models of the evolution of supermassive black hole (SMBH) pairs in post-merger remnant galaxies are necessary to motivate
observational searches for dual active galactic nuclei (AGN) and gravitational wave sources. Studies have explored the dynamical evolution of SMBH pairs under the influence of dynamical friction to calculate pairing times and predict the expected population of dual-AGNs at various redshifts.  We formulate a three-dimensional dynamical model of SMBH pairs in the innermost kiloparsec of a post-merger galaxy to investigate the impact of orbital inclination with respect to the galactic disk on pairing times. The SMBH pairs are evolved in 81 different galaxy configurations initialized using a Gauss-Seidel Poisson solver. The dynamics are calculated for 12 distinct initial inclinations ranging from 0 to 75 degrees in each of the galaxies to gauge the impact of inclination on pairing time. Orbits characterized by initial inclinations greater than 20 degrees frequently require longer pairing times when compared to uninclined orbits. Pairing times for orbits with inclinations $\gtrsim 45$ degrees often exceed 14 Gyr. Galaxies with higher mass SMBH pairs and faster rotating disks generally shorten pairing times relative to galaxies with less massive or slower rotating disks when the inclination is $\lesssim 45$ degrees. The model suggests that SMBH pairs that form from mergers at inclinations $\lesssim 20$ degrees are likely progenitors of dual-AGN and gravitational wave sources.

\end{abstract}


\section{Introduction}
\label{sec:intro}
A fundamental prediction of the hierarchical model of galaxy formation is the growth of massive galaxies via mergers \citep[e.g.][]{1978MNRAS.183..341W, 1991ApJ...379...52W}. Since supermassive black holes (SMBHs) with masses between $10^6 - 10^{10} M_{\rm \odot}$ are expected to reside in the centers of most massive galaxies \citep[e.g.][]{1982MNRAS.200..115S, 1995ARA&A..33..581K, 1998AJ....115.2285M,  2005SSRv..116..523F}, it is expected that merger galaxies will frequently contain two SMBHs that, in some cases, shine as dual active galactic nuclei (dAGNs) due to the significant nuclear gas inflows resulting from the merger \citep[e.g.][]{1991ApJ...370L..65B, 1999PNAS...96.4749F, Hopkins_2005}. Simulation studies such as those by \cite{2017MNRAS.469.4437C} and \cite{10.1093/mnras/stac1217} have quantified the incidence and observability of dAGNs across merger stages, finding that activity persists for tens to hundreds of Myr and often precedes SMBH coalescence. Therefore, identifying a population of dAGNs in post-merger galaxies is crucial to verifying the current paradigm of galaxy formation and evolution. Theoretical models of dAGN populations are vital to upcoming observational surveys such as the Next Generation Very Large Array \citep[ngVLA;][]{McKinnon_2016} in the radio band and newAthena \citep[][]{2025NatAs...9...36C} in the X-ray band whose scientific goals include the detection of dAGNs.

In the framework for binary SMBH formation described by \cite{BBR1980}, the physical mechanisms driving the formation of the binary vary significantly depending on the separation between the two SMBHs. Following a galaxy merger, when the separation between the two SMBHs is $\sim 1$ kpc, the principal driver of the orbital decay of the SMBH pair is dynamical friction (DF). DF gradually drains the orbital energy of the pair through the formation of an overdense wake behind the orbiting SMBHs. This wake is generated from ambient gas \citep[][]{O1999, KK2007} and stars \citep[][]{C1943, AM2012} in the post-merger galaxy. Once the separation reaches $\sim 1$ pc, the engine of the decay becomes stellar ``loss-cone" scattering \citep[][]{Q1996, 2002MNRAS.331..935Y} which drains the energy of the pair via the ejection of stars. After the pair reaches a separation of $\sim 1000$ Schwarzschild radii, gravitational wave emission begins to dominate the orbital decay until the binary coalesces \citep[][]{KT1976, 1976ApJ...204L...1T, LISA2017}. Therefore, identifying dAGNs provides a promising avenue for cataloging future SMBH mergers and GW events. 

An important aspect of the dynamics neglected in many studies of black hole pair evolution at kpc scales is the inclination of the orbiting SMBH relative to the galactic disk. This is important to consider since the efficiency of dynamical friction is dependent on the background density of gas and stars which, for an inclined orbit, will vary substantially as the SMBH oscillates through the plane of the disk. Earlier studies have investigated the SMBH pairing probability and timescale using semianalytic orbital evolution models \citep[e.g.,][]{2012MNRAS.423.2533B, 2017MNRAS.464.2952T,Tremmel_2018, LBB20b, LBB20a}. In particular, \citet[][hereafter Li20A]{LBB20b} analyze the evolution of SMBH pairs in a merger remnant disk galaxy under the influences of stellar and gaseous DF within the plane of the disk and construct the distribution of pairing times for a suite of orbits in various galactic environments. Cosmological simulations demonstrate that neighboring galaxies orbit each other at a wide range of spin-orbit angles \citep[e.g.][]{Moon_2021}. It is therefore pertinent to quantify the extent to which the relative inclination of the merging galaxies affects the pairing time of an orbiting SMBH pair resulting from such a merger.

In this paper, we analyze the effects of orbital inclination on an SMBH pair using a 3D dynamical computation to integrate the equations of motion governing the evolution of the SMBH pair. A refined-mesh Poisson solver is formulated to compute the gravitational potential acting on the SMBH pair and evaluate the effect of gaseous and stellar DF. The orbital dynamics are computed over a suite of approximately 1,000 different black hole pairs in post-merger galaxies governed by a similar set of parameters used by Li20A. A distinct parameter $i_0$ is introduced, specifying the angle by which the two SMBHs are initially separated relative to the plane of the galactic disk. The principal goal of this study is to demonstrate the influence of this parameter on the pairing time, which is defined to be the time taken by the smaller, secondary SMBH to reach a separation of 10 pc from the more massive primary at the center of the galaxy. 

The rest of this paper is structured as follows. In Section \ref{sec:methods}, the model for a merger galaxy is introduced and the calculation of the gravitational potentials and DF is described. The setup of the computation and the procedure for calculating the pairing times are also established. In Section \ref{impact}, the impact of inclined orbits is thoroughly explored for a few fiducial test cases to predict how pairing times vary with increasing inclination. The full suite of parameters is analyzed in Section \ref{Analysis} where we probe how changing the inclination in conjunction with different facets of the galaxy's structure alters the dynamics of pairing. We discuss the implications of these calculations in Section \ref{sec:discuss} and conclude in Section \ref{sec:concl}.
\section{Methods}
\label{sec:methods}
\subsection{Modeling the Centers of Merger Remnant Galaxies with Numerical Poisson Solvers}
The axisymmetric merger remnant galaxy is modeled similarly to the Milky Way using three distinct elements: a stellar disk, a gas disk, and a stellar bulge. We omit dark matter halos from our model since their contribution to the galactic potential within $\sim 1$ kpc is small relative to the other components (see Section \ref{simplifying}). In all calculations,  a cylindrical coordinate system is used, and the more massive primary SMBH is fixed at the center of the galaxy ($r = z = 0$). This approximation is suitable provided the mass ratio between the secondary and the primary is $\lesssim 0.1$ so that the displacement of the primary due to the secondary’s gravitational influence is negligible compared to the secondary’s orbital radius. Each structure is defined by a density distribution which is used to compute the gravitational potential at each point in space and evaluate the strength of DF during the evolution of the SMBH pair. The density distributions for the stellar disk, gas disk, and bulge are carried over from Li20A and generalized to a 3D environment. The parameters utilized to determine the structure of the galaxy are the mass of the central SMBH $M_1$, the central gas density $n_{\rm gd}$, the gas fraction of the galaxy $f_{\rm g}$, and its rotational speed $v_{\rm g}$, defined as a constant fraction of the local circular speed of the disk, $v_c(r,z)$. The gas fraction is defined to be the ratio of the gas disk mass to the total mass of both the gas and stellar disks, which does not include the stellar bulge. The mass of the secondary is permanently set to be 1/9 the mass of the primary to avoid straining the assumption that the central SMBH stays fixed. This choice is also motivated by cosmological models and simulations which suggest that SMBH mergers most commonly occur at unequal mass ratios on the order of $1/10$ \citep[e.g.,][]{2007MNRAS.377.1711S, 2009ApJ...697.1621G, 2022MNRAS.510..531C}.

For the stellar bulge, we use the model described by \cite{BT1987},
\begin{equation}\label{rhob}
    \rho_{\rm b}(r,z) = \rho_{\rm b0} \left( \frac{m}{a_{\rm b}}\right)^{-\alpha_{\rm b}}e^{-m^2 / r_{\rm b}^2}
\end{equation}
where $m \equiv \sqrt{r^2 + z^2/q_b^2}$.  The coefficients in Equation \ref{rhob} are set to be $a_{\rm b} = 1$ kpc, $r_{\rm b} = 1.9$ kpc, $\alpha_{\rm b} = -1.8$, and $q_{\rm b} = 0.6$ \citep[][]{BT1987}. The normalization constant $\rho_{\rm b0}$ is calculated such that the total integrated mass within 2 kpc is 1000 $M_1$ \citep{1998AJ....115.2285M}. Note that manipulating $M_1$ alters the entire structure of the galaxy as a result of this relation. 

The density distributions of the stellar and gas disks both follow double exponential profiles. For the stellar disk,
\begin{equation}\label{rhos}
    \rho_{\rm s}(r,z) = \Sigma_{\rm d} \, e^{-r/R_{\rm d}} \left(\frac{1}{4z_{\rm 0}}e^{-|z|/z_{\rm thin}} +  \frac{1}{4z_{\rm 1}}e^{-|z|/z_{\rm thick}} \right)
\end{equation}
where $R_{\rm d}$ is the characteristic radius of the disk, $z_{\rm thin}$ is the characteristic height of the thin disk, $z_{\rm thick}$ is the characteristic height of the thick disk, and $\Sigma_{\rm d}$ is the central surface density. The characteristic radius $R_{\rm d}$ is determined by the mass of the central SMBH through the relation $R_{\rm d} = \log{(M_1 / 10^5 M_{\rm \odot})}$ kpc. The heights are then specified by $z_{\rm thin} = 0.1R_{\rm d}$ and $z_{\rm thick} = R_{\rm d}/3$. The normalization constant $\Sigma_{\rm d}$ is chosen so that the galaxy maintains the appropriate gas fraction $f_{\rm g}$ which is specified in advance and in this study takes a range of values between 0.3 and 0.7 (Table \ref{tab:params}). The density distribution of the gas disk is calculated as
\begin{equation} \label{rhog}
    \rho_{\rm g}(r,z) = n_{\rm gd} \, m_{\rm p} \, e^{-(r/R_{\rm g} + |z| / z_{\rm g})}
\end{equation}
where $n_{\rm gd}$ is the central number density of gas particles, $m_{\rm p}$ is the mass of the proton, $R_{\rm g}$ is the characteristic radius of the gas disk, and $z_{\rm g}$ is its characteristic height. The term $n_{\rm gd}$ is chosen to take values between 100 and 300 cm$^{-3}$ (Table \ref{tab:params}) and $R_{\rm g}$ is given by $R_{\rm g} = 2R_{\rm d}$. The height of the gas disk is set to be $z_{\rm g} = 100$ pc to be in approximate alignment with the thickness of the Milky Way \citep{BT1987, 2023Galax..11...77V}. 

With these density profiles, the potential, $\Phi$, may be computed according to the Poisson equation in cylindrical coordinates 
\begin{equation}
    \frac{1}{r} \frac{\partial}{\partial r} \left( r \frac{\partial \Phi}{\partial r}\right) + \frac{\partial^2 \Phi}{\partial z^2} = 4\pi G \, \rho(r,z)
\end{equation}
where in this case $\rho(r,z)$ would be the sum of all the density distributions given in equations \ref{rhob}, \ref{rhos}, and \ref{rhog}. Note that there is no dependence on the azimuthal angle $\phi$ due to the assumption of axisymmetry. Only vacuum boundary conditions are considered, i.e., $\Phi(\infty) \rightarrow 0$. In principle, this could be solved exactly using Green's functions, but it would require cumbersome numerical integration at every time step in the orbit. Under the assumption that the mass distribution is static at all times, this can be remedied by using a grid-based numerical Poisson solver that calculates the potential before running the dynamical solver for the orbit of the SMBH pair.

\begin{figure}
    \centering
    \includegraphics[width=1.0\linewidth]{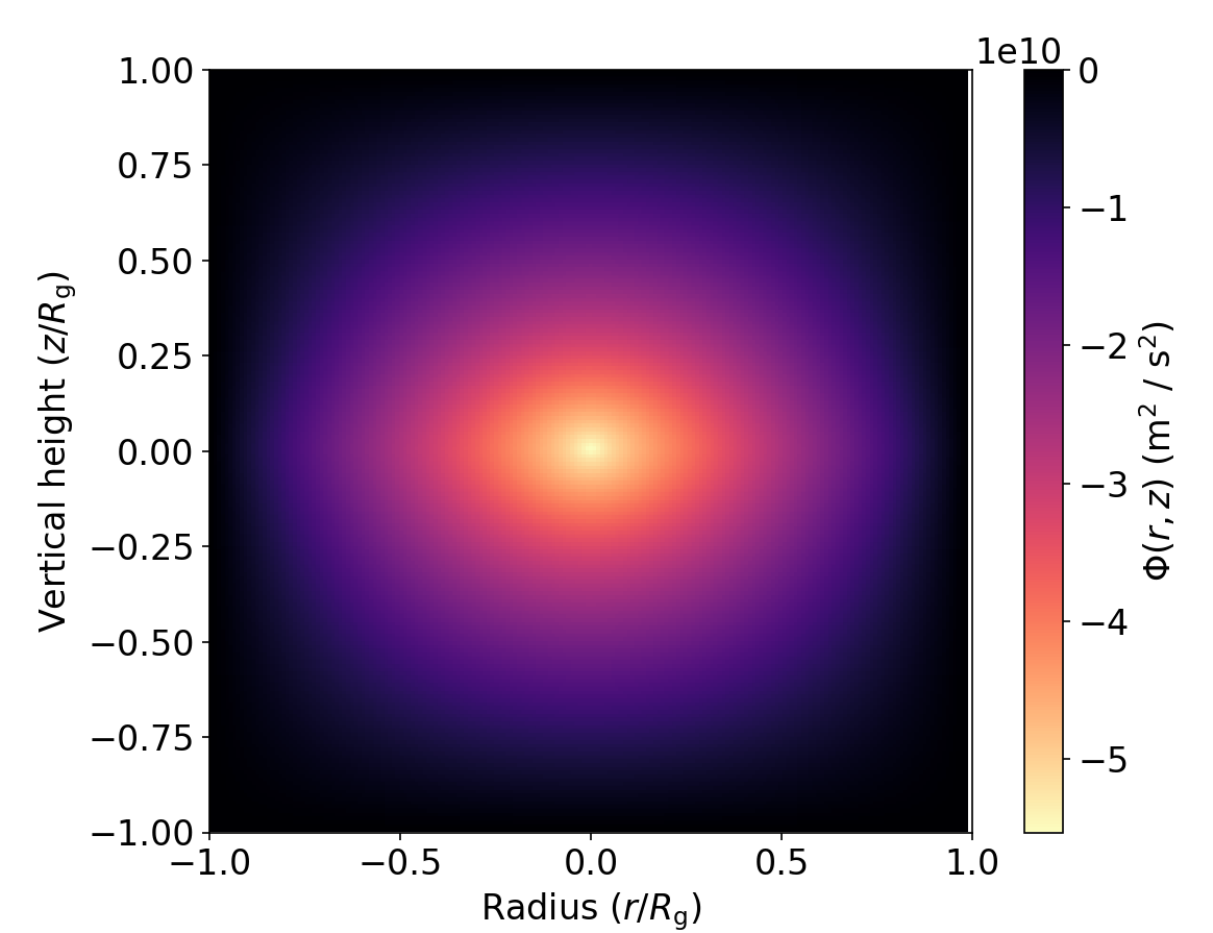}
    \caption{Illustration of a gravitational potential field computed from the specified mass density profiles for the galactic bulge, stellar disk, and gas disk. In this scenario, the mass of the central SMBH is $M_1 = 10^6$ \sm, the central gas number density is $n_{\rm \rm gd} = 100$ \cm, and the gas fraction is $f = 0.7$.}
    \label{fig:1}
\end{figure}

\begin{deluxetable*}{ccC}
\tablenum{1}
\tablecaption{Galaxy Model Parameters\label{tab:params}}
\tablewidth{0pt}
\tablehead{
\colhead{Symbol} & \colhead{Description} & \colhead{Values}
}
\startdata
$M_1$\ & central MBH mass ($M_{\rm \odot}$) &  10^6, 10^7, 10^8 \\
 \rhog\ & central gas number density ({\rm cm}$^{-3}$) & 100, 200, 300 \\
 $f_{\rm g}$\ &  gas disk mass fraction & 0.3, 0.5, 0.7 \\
$v_{\rm \rm g}$ \ & gas disk rotational speed
 & 0.3, 0.5, 0.7 \\
 $i_0$\ & initial orbit inclination angle (deg) & 0.0, 2.5, 5.0, 7.5 \\ & & 10.0, 15.0, 20.0, 25.0 \\ & & 30.0, 45.0, 60.0, 75.0
\enddata
\tablecomments{The values of the disk speed are given in units of the local circular velocity $v_c(r,z)$. For example, a disk speed of 0.3 corresponds to a disk rotating at 0.3 times the local circular velocity at each point in the disk. Also note that $M_2$ is set to be $M_1/9$ in every calculation.}
\end{deluxetable*}

To formulate the Poisson solver, an $N \times M$ grid is initialized to discretize the space. Each cell is populated with a density value given by the density profiles previously specified. From this, the solver computes the gravitational potential corresponding to each cell using Gauss-Seidel relaxation \citep[e.g.][]{press2007}. During orbit integration, the potential at the location of the secondary SMBH is evaluated by interpolating between these cells. A gravitational potential computation is provided for an example galaxy in Figure \ref{fig:1}. In this example, the solver utilizes a square grid with resolution of $N = M = 200$. The side length of the square grid is set to $2R_{g}$, so the resolution elements are squares with side length $0.01R_g$, which equals 20 pc in the galaxy from Figure \ref{fig:1}. The grid is constructed to be relatively coarse since the computations test a large collection of parameters, each specifying a new galaxy. However, the calculation necessitates enhanced accuracy toward the midplane of the galaxy at $z = 0$ due to the steeper density gradients near the midplane. For this reason, another refined mesh is employed covering the central region of $-0.05R_{\rm g} < z < 0.05R_{\rm g}$. This inner region is initialized in the same way as the previous grid, except the boundary conditions are altered so that the potential of the inner region matches the potential of the outer region at the boundary. The resolution of the inner region is again set to $N = M = 200$ to maintain adequate accuracy for coplanar orbits. The resolution elements are rectangles with width $0.01R_g$ and height $5 \times 10^{-4} R_g$. Since $R_g$ varies between 2 kpc and 6 kpc, the heights of the inner cells vary between 1 and 3 pc. The calculation is robust to changes in the spatial resolution\footnote[1]{We randomly sample 10 computations and test this robustness by both doubling and halving the resolution in each trial. In all cases, the corresponding pairing times change by no more than 10\%.}. The error tolerance used as the convergence condition for this algorithm is specified by $\epsilon_{\rm out} =2.0\times 10^{-6} \Phi_{\rm \rm \epsilon}$ for the outer region and by $\epsilon_{\rm in} = 1.0 \times 10^{-7} \Phi_{\rm \rm \epsilon}$ for the inner region where $\Phi_{\rm \rm \epsilon} \equiv 4\pi G n_{\rm gd} m_{\rm p} R_{\rm g}^2$. The galaxy in Figure \ref{fig:1} serves as a fiducial example of a configuration that we refer to in the rest of the paper.

\subsection{Computing the Orbital Dynamics}

The SMBH pair is evolved using a fourth-order Runge-Kutta solver to numerically solve the equations of motion governing the secondary SMBH in the $r$, $\phi$, and $z$-coordinates. We calculate the equations of motion for $r$ and $\phi$ in the same manner as Li20A. The equation of motion for $z$ is calculated independently using the methods outlined in this section. An adaptive time step is utilized and set to at most 1\% of the instantaneous circular period of the SMBH, which maintains conservation of energy and angular momentum to within 1\%. The secondary is assumed to be ``bare" such that it is stripped of all the stars, gas, and dark matter that might have been bound to it in the early stages of the galaxy merger. 

The time-evolution of the secondary SMBH is governed by the gravitational potential of the galaxy and the force of dynamical friction from ambient stars and gas. The net gravitational force acting on the SMBH is calculated by evaluating the potential using the interpolation functions \texttt{gsl\_interp\_2d\_{\rm d}eriv\_x} and \texttt{gsl\_interp\_2d\_{\rm d}eriv\_y} from the GNU scientific library \citep{gsl}. The functions take in an array representing the potential at each cell of the computation space along with the $r$ and $z$ coordinates of the secondary to output the gravitational acceleration in the $r$ and $z$ directions. The calculation of DF is identical to the treatment described in Li20A. Stellar and gaseous DF are computed using the models described by \cite{AM2012} and \cite{KK2007}. Stellar DF is calculated according to Equation 30 in the paper by \cite{AM2012}. This equation directly relates the DF force vector and the velocity vector, so the $z$-component of the stellar DF force is calculated using the $z$-component of the velocity. Incorporating gas DF requires a treatment for the sound speed $c_{\rm s}$ of the gas disk, which is calculated using a modified version of the Toomre stability criterion \citep[][]{1964ApJ...139.1217T}. The modified criterion is given by 

\begin{equation}
    c_s(r,z) = \frac{\pi G rz_g\rho_{\rm g}(r,z)}{v_g(r,z)}
\end{equation}
where $v_g(r,z)$ is the local gas disk rotation speed. Our model replaces the surface mass density in the Toomre model with the function $2\rho_{\rm g}(r,z) z_{\rm g}$. This approximately corresponds to the surface density of the nonuniform gas disk. With this model, the sound speed distribution varies in space and is calculated to be on the order of 10 km/s. Temperatures corresponding to this sound speed are on the order of $10^4$ K, which is expected for the turbulent environment of a post-merger galaxy \citep[][]{1991ApJ...370L..65B}.

\begin{figure*}
    \centering
    \includegraphics[width=1.0\textwidth]{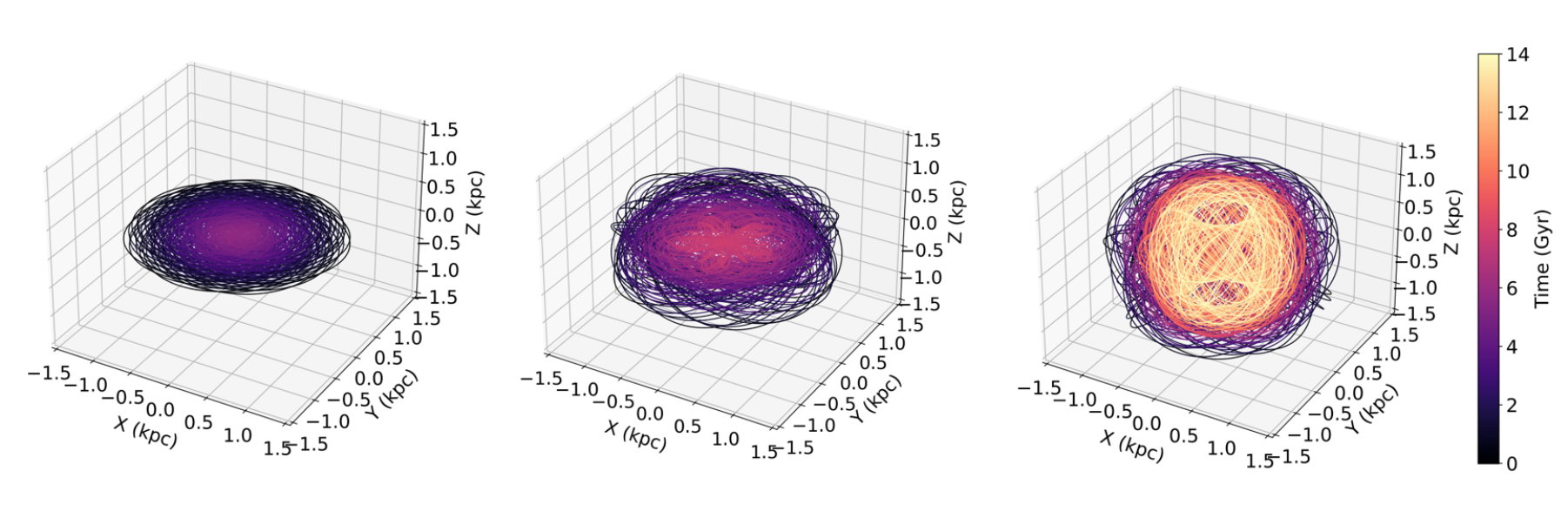}
    \caption{Example orbits in a galaxy with $n_{\rm gd} = 100$ \cm, $f_{\rm g}$ = 0.7, $v_{\rm g} = 0.7v_c$, and $M_1 = 10^6$ \sm. \textit{Left:} An orbit at an initial inclination of 0$^{\circ}$ which merges in 5.75 Gyr staying at the same inclination of 0$^{\circ}$ for the whole orbit. \textit{Middle:} An orbit at an initial inclination of 25$^{\circ}$ which merges in 7.90 Gyr. The orbital inclination of the secondary decreases, and  the SMBH settles into the plane of the disk due to vertical DF. \textit{Right:} An SMBH with initial inclination of 45$^{\circ}$, which does not pair within a Hubble time. The secondary maintains a large inclination throughout the entire orbit since vertical DF is inefficient far from the plane of the disk when $v_z \gg c_s$.}
    \label{fig:2}
\end{figure*}
Previous calculations of gas DF have relied on the assumption that the perturbing mass travels through an infinite, homogeneous medium \citep[][]{O1999, KK2007}. Since the SMBH potentially exits the galactic disk on inclined orbits, the density of the medium varies substantially during oscillations through the disk plane, so there is no analytic expression for gas DF in the $z$-direction. We introduce a model to approximately incorporate the effects of gas DF on the $z$-dynamics. This effect will be referred to as vertical DF in the remainder of the paper. The Mach number experienced by the secondary SMBH moving vertically through the gas disk (hereafter ``vertical Mach number" for brevity) is defined as $\mathcal{M}_{\rm z} \equiv |\dot{z}| / c_{\rm s}$ which dictates the strength of vertical gaseous DF. In this model, we employ the expression for gas DF calculated by \cite{O1999} which is a piecewise function of the vertical Mach number $\mathcal{M_{\rm z}}$,
\begin{equation} \label{gasdf}
    \vec{F}_{\rm g} = -\frac{4\pi (GM_2)^2 \rho_{\rm g}(r,z)}{v_z^2} I(\mathcal{M}_{\rm z}),
\end{equation}
where $\rho_{\rm g}$ is the local gas density and $v_z$ is the vertical component of the speed of the secondary relative to the gas. The dimensionless coefficient $I(\mathcal{M}_{\rm z})$ varies for the subsonic and supersonic cases,
\begin{equation}
    I(\mathcal{M}_{\rm z}) = \begin{cases}
      \frac{1}{2}\ln{\left(\frac{1+\mathcal{M}_{\rm z}}{1 - \mathcal{M}_{\rm z}} \right)} - \mathcal{M}_{\rm z}, \ \mathcal{M}_{\rm z} < 1 \\ \frac{1}{2} \ln\left(1 - \frac{1}{\mathcal{M}_{\rm z}^2}\right) + \ln\left(\frac{Vt}{r_{\rm \min}}\right), \ \mathcal{M}_{\rm z} > 1
       \end{cases}
\end{equation}
 The term $\ln\left( {Vt} / {r_{\rm min}}\right)$ is set equal to 10 to obtain a force curve similar to the one computed in Figure 3 of the study by \cite{O1999}. Note that this function suffers from a singularity at $\mathcal{M}_{\rm z} = 1$, so the $I(\mathcal{M}_z)$ coefficient is set to $I(\mathcal{M}_{\rm z}) = I(0.999)$ in the small region of $0.999 \leq \mathcal{M}_{\rm z} \leq 1$. An important edge case that poses numerical issues is $v_z$ going to 0, which happens at the turning points of the vertical oscillations. In such a case, Equation \ref{gasdf} diverges due to division by 0. To remedy this, for Mach numbers less than 0.1, the subsonic expression for $\vec{F_{\rm g}}$ is replaced with its first order Taylor expansion so that the drag force is linear. This avoids the problem of small divisors while maintaining an accurate and physically motivated model for gas DF.

Each calculation is governed by five parameters specifying the structure of the galaxy and the nature of the orbit: $M_1$, $n_{\rm gd}$, $f_{\rm g}$, $v_{\rm g}$, and $i_0$, the initial inclination angle of the secondary relative to the disk plane. The set of parameters explored is enumerated in Table \ref{tab:params}. The range of $f_g$ is consistent with the high gas-fractions observed at $z \gtrsim 2$ when galaxy mergers are more frequent \citep[][]{2009ApJ...702..307S, 2010Natur.463..781T, 2010gfe..book.....M, 2017MNRAS.464.2952T}. The initial inclination $i_0$ is defined by the relation $\tan{i_0} = z_0 / r_0$ where $z_0$ is the initial height of the secondary relative to the plane of the disk and $r_0$ is its planar radius.

The calculation encompasses a particularly large set of values for $i_0$ compared to the other parameters since this analysis focuses mostly on analyzing the effects of inclination on the pairing time. A non-uniformly spaced set of inclinations is adopted to better reflect the distribution of spin-orbit angles by \cite{Moon_2021}. Analysis from this study indicates that galaxy pairs are most likely to undergo prograde orbits with aligned rotation axes, which is what motivates the choice of parameters. Since the decay of the SMBH pair is largely driven by dynamical friction in the galactic disk, tilting the orbit away from the plane of the disk is expected to increase the pairing time since the strength of DF depends on the local density of gas and stars. From equations \ref{rhos} and \ref{rhog}, the density distribution decays exponentially in the radial and vertical directions. However, since the scale heights are smaller than the scale radii, the density decays faster in the vertical direction, which implies that orbits tilted away from the plane of the disk will generally lose energy from DF at a slower rate. Therefore, the pairing time is expected to \textit{increase} with inclination, implying that the pairing times found in Li20A may serve as a lower bound for the special case of zero inclination.

In each computation, the secondary is placed at a distance of 1 kpc away from the center such that $\sqrt{r_0^2 + z_0^2} = $ 1 kpc.\footnote{This distance is selected in order to sufficiently resolve the evolution at separations where DF is the dominant decay mechanism \citep[][]{ 2022LRR....25....3B}. This distance scale also complements the results of cosmological simulations, which typically follow the evolution of SMBH pairs down to $\sim 1$ kpc \citep[e.g.][]{2014MNRAS.445..175G, 2017MNRAS.464.3131K}.} The speed of the secondary at the beginning of each orbit is always set to be $v_{\rm 1kpc}(1+f)^{1/2}$ where $v_{\rm 1kpc}$ is the circular velocity at a radius of 1 kpc and a height of 0 kpc and $f$ is set to 0.5 (Li20A). The secondary is initialized with its velocity in the $+\hat{\phi}$ direction following the rotation of the disk so that all orbits are prograde. Each computation terminates if the secondary reaches the central 10 pc of the galaxy or if 14 Gyrs elapse, indicating no pairing within a Hubble time. We refer to the time taken to reach the inner 10 pc of the galaxy as the pairing time $t_{\rm p}$ throughout the rest of the paper. The pairing time of each orbit is evaluated as a function of all five parameters. 

\section{Impact of Orbital Inclination on Inspiral Time for a Fiducial Galaxy}\label{impact}

\begin{figure*}
    \centering
    \includegraphics[width=1.0\textwidth]{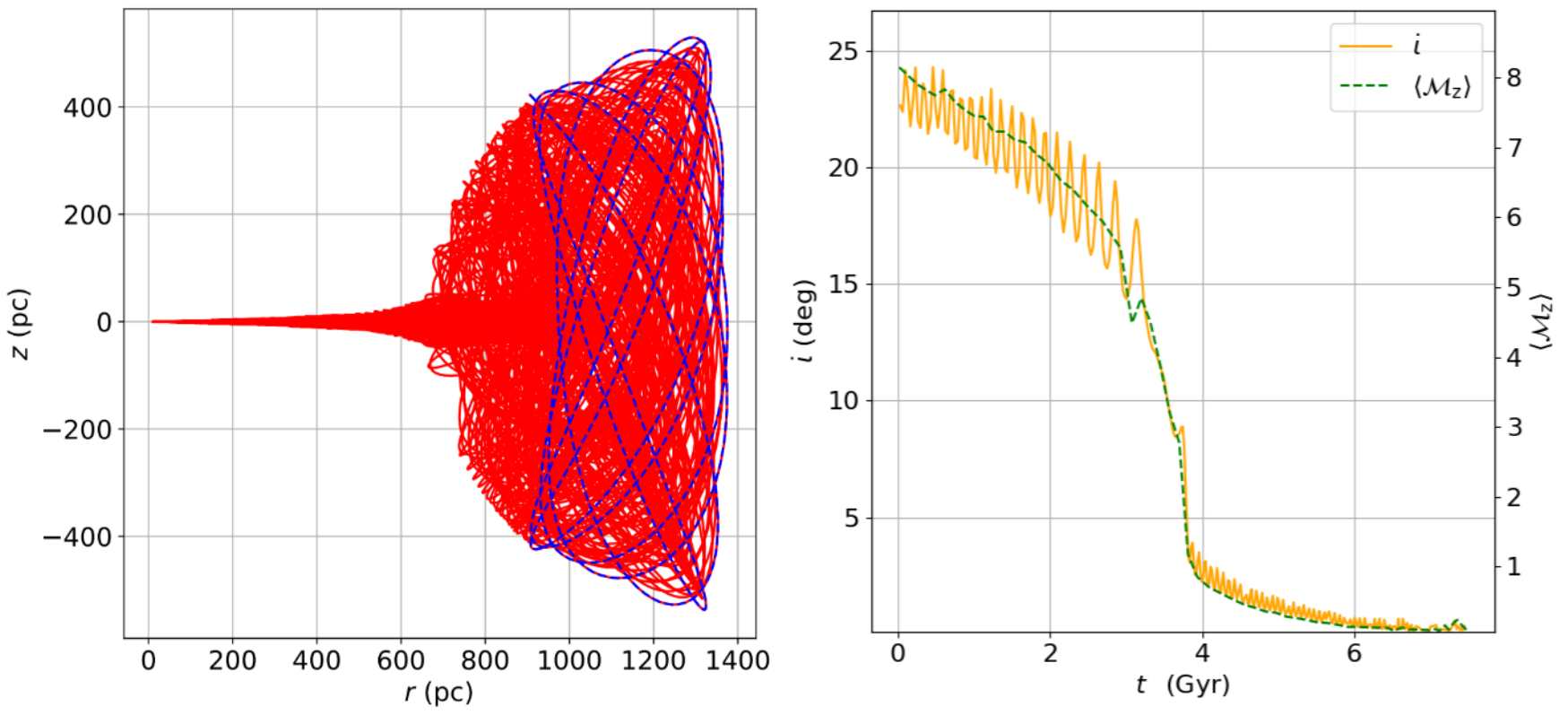}
    \caption{\textit{Left:} Projection of the full 25 degree orbit from the middle panel of Figure \ref{fig:2} into the $rz$-plane represented by the solid red curve. The amplitude of vertical oscillations abruptly decays when the vertical Mach number is of order unity, leading to the very thin envelope observed at $r \lesssim 600$ pc. The dashed blue line indicates the shape of the orbit for the first 200 Myr when DF has not substantially altered the dynamics. \textit{Right:} Orbital inclination as a function of time for the same orbit plotted as the solid orange curve. The vertical Mach number is plotted as the dashed green curve alongside the inclination to demonstrate the strong relationship between vertical speed and vertical DF strength. At a critical time around 4 Gyr, vertical DF becomes extremely efficient and drags the secondary close to the plane where it maintains a low inclination for the rest of the orbit. This coincides with the vertical Mach number decreasing to the range where $\langle \mathcal{M}_z\rangle \sim 1$.}
    \label{25inc}
\end{figure*}

\begin{figure*}
    \centering
    \includegraphics[width=1.0\textwidth]{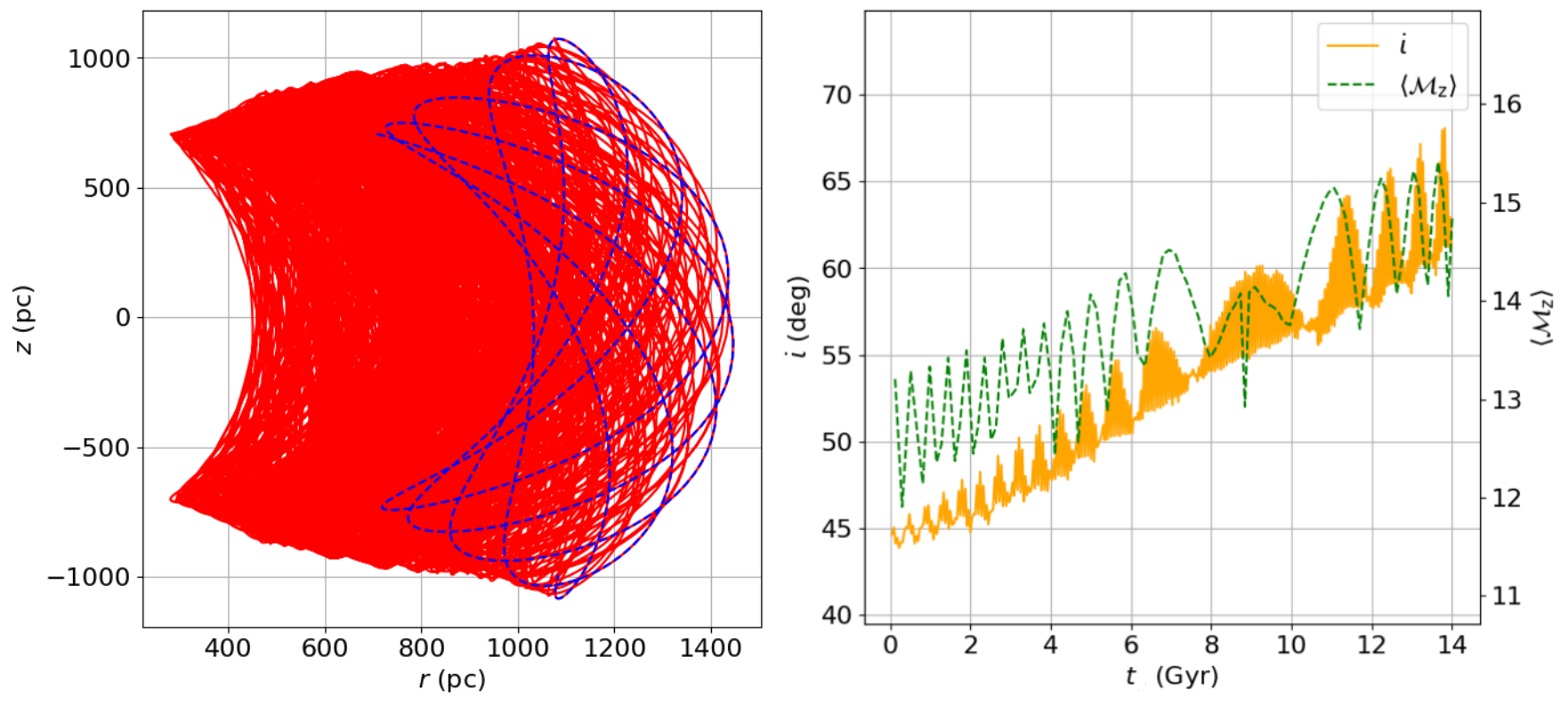}
    \caption{\textit{Left:}  Projection of the full 45 degree orbit from the right panel of Figure \ref{fig:2} into the $rz$-plane. The amplitude of vertical oscillations decays more gradually compared to Figure \ref{25inc} since the vertical Mach number is always much greater than 1. The envelope gradually shifts due to DF, but does not collapse into a thin line as in the 25 degree case. \textit{Right:} Orbit inclination as a function of time for the 45$^{\circ}$ inclined orbit. In contrast to the behavior observed in the right panel of Figure \ref{25inc}, the inclination $i$ \textit{increases} on average. The inclination increases due to the azimuthal DF being more efficient than vertical DF, which drains the $z$-component of the angular momentum faster than the other components of the angular momentum.}
    \label{45inc}
\end{figure*}

The goal of this section is to describe the physical processes that influence the pairing times of individual SMBH pair configurations. Three examples are explored in detail to illustrate how the distinct parameters governing the structure of a galaxy dictate the pairing probability and pairing time of the SMBHs (Figure \ref{fig:2}). The galactic structure is the same as the galaxy described in Figure \ref{fig:1}. The secondary is launched in the azimuthal direction at a distance of 1 kpc away from the central primary in each trial, but $i_0$ is varied between 0\dg, 25\dg, and 45$^{\circ}$ corresponding to the left, middle, and right panels of Figure \ref{fig:2} respectively. The speed of the secondary is set to be the same in each trial, so less inclined orbits begin the computation with less total energy due to the increased gravitational potential of the disk. All three trials are initiated with the same total angular momentum.

The 0$^{\circ}$ inclination orbit in the left panel of Figure \ref{fig:2} is utilized as a control. This orbit decays in 5.75 Gyr primarily due to gas DF and stellar bulge DF which dominate due to the Mach number of the secondary oscillating around $\mathcal{M} \sim 1$ and the secondary effectively orbiting inside the bulge during the latter part of the orbit. When $\mathcal{M} \sim 1$, gas DF is most efficient, which leads to shorter pairing times. When $i_0$ is increased to 25$^{\circ}$ in the middle panel of Figure \ref{fig:2}, a significant change in the dynamics occurs. In particular, $t_p$ increases by more than 2 Gyr. Further increasing the initial inclination to 45${^\circ}$ in the right panel of Figure \ref{fig:2} results in no pairing at all. It is evident that $t_p$ increases significantly with inclination angle, reaching a point where pairing is impossible within 14 billion years at some $i_0$ between 25 and 45 degrees.

The shapes of the orbits indicate the mechanisms responsible for this disparity in the pairing times. The left panel of Figure \ref{25inc} displays a projection of the 25 degree orbit from the second panel of Figure \ref{fig:2} into the $rz$-plane where the spatial evolution of the orbit is more apparent. The secondary orbits above the $z=0$ plane for approximately 4 Gyr before returning to a nearly coplanar orbit. The inclination of the secondary decreases with radius due to vertical DF gradually dragging the secondary into the plane of the disk where DF is generally more efficient. To quantify this dragging effect directly, the evolution of the instantaneous orbit inclination $i$ is plotted as a function of time for the 25$^{\circ}$ case in the right panel of Figure \ref{25inc}. The instantaneous inclination is defined as $i \equiv \arctan{[z_a(t) / r_a(t)]}$ where $z_a(t)$ is the amplitude of vertical oscillations (i.e., the height of the turning point) as a function of time and $r_a(t)$ is the instantaneous radius of the secondary measured at each peak of the vertical oscillations. The inclination oscillates due to the varying eccentricity of the orbit and gradually decays for about 3 Gyr. The secondary SMBH then undergoes a sudden drop in inclination until about 4 Gyr. From then on, the orbital inclination steadily decays to 0 degrees until the secondary undergoes pairing. The inclination evolution can be explained in terms of how vertical DF scales with the average vertical Mach number designated by the green dashed line in the right panel of Figure \ref{25inc}. For roughly the first 3 Gyr, the vertical Mach number averaged over each period, $\langle \mathcal{M}_{\rm z} \rangle$, is high ($5 \lesssim \langle \mathcal{M}_{\rm z} \rangle \lesssim 8$). Therefore, vertical DF can only drain the kinetic energy and the $z$-component of the momentum at a low rate. This leads to a steady decrease of the average vertical Mach number until it is of order unity, at which point vertical DF is very efficient. This causes the sudden drop in inclination at 4 Gyr. As the $z$-component of the speed of the secondary continues to diminish, the Mach number again reaches a regime where vertical DF is inefficient, so the inclination decreases very gradually until the pairing occurs.

The dramatic change in dynamics at larger inclinations is apparent for the 45$^{\circ}$ orbit corresponding to the right panel of Figure \ref{fig:2} and the left panel of Figure \ref{45inc} where the secondary maintains a large inclination throughout the entire orbit. Vertical DF is unable to drag the secondary into the galactic disk since it is exposed to very little background matter. This is compounded by the fact that vertical DF becomes less efficient with increasing $\mathcal{M}_{\rm z}$ (see Equation \ref{gasdf}). Since the local sound speed $c_{\rm s}$ is lower far away from the disk and the vertical oscillations are faster for larger inclinations, DF becomes extremely inefficient. Therefore, at large inclinations, pairing within a Hubble time becomes unlikely.

The inclination in the 45$^{\circ}$ case undergoes an average increase over the span of the orbit as shown in the right panel of Figure \ref{45inc}. This result can again be understood by considering the evolution of the average vertical Mach number (Figure \ref{45inc} [right]; dotted line). At large $i$, it is possible that vertical DF never becomes sufficiently strong to drag the secondary back into the plane of the disk. In this case, azimuthal DF may drain the $z$-component of the angular momentum at a higher rate than the vertical DF can reduce the other components of angular momentum. Therefore, the orbit of the secondary can become more tilted even if the vertical amplitude of oscillations is decreasing. This is the phenomenon causing the observed increase in inclination in Figure \ref{45inc}. 

\begin{figure}
    \centering
    \includegraphics[width=1.0\linewidth]{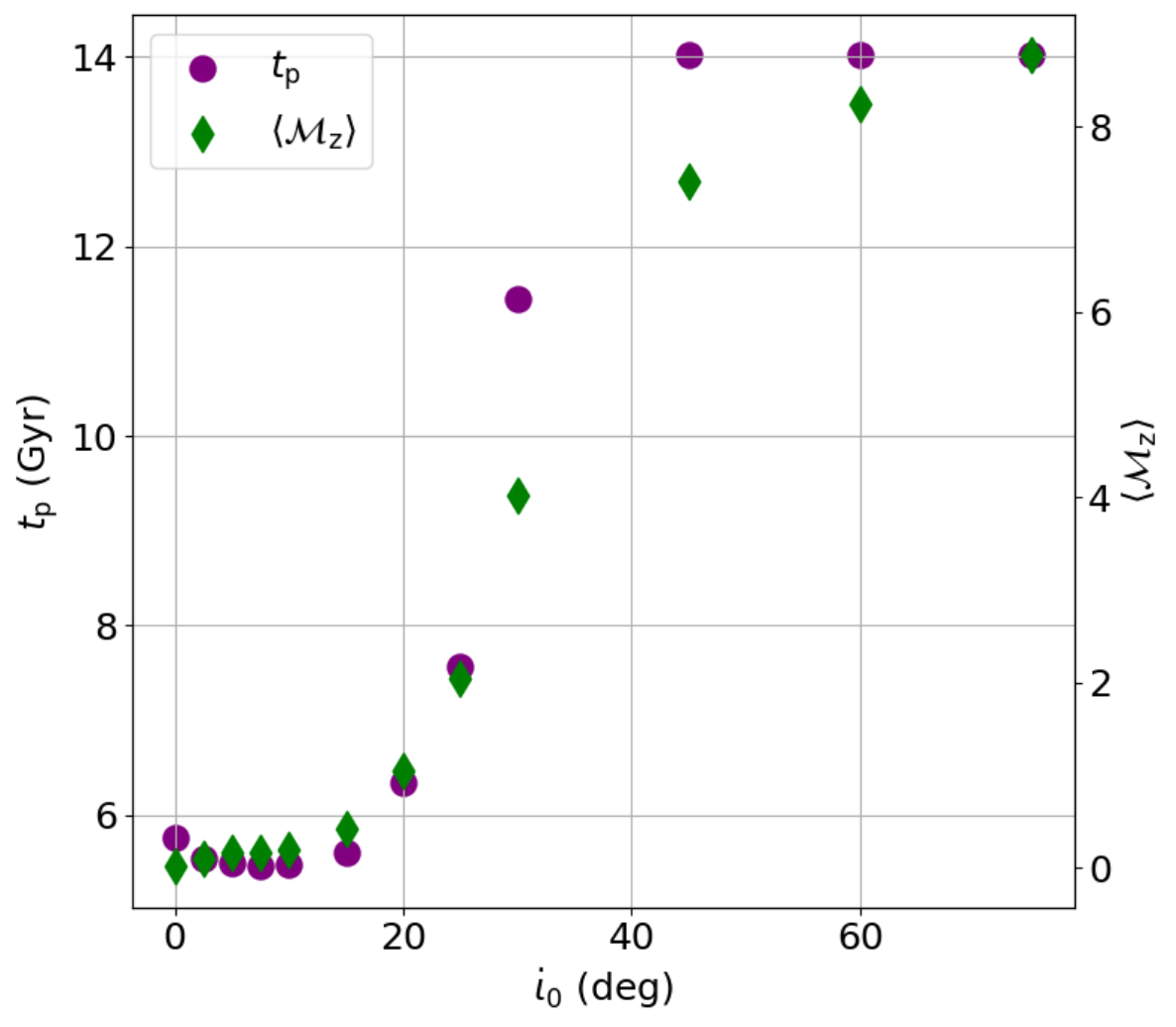}
    \caption{\textit{Left axis:} Pairing time (purple circles) as a function of initial inclination angle $i_0$ for orbits in the same galaxy displayed in Figure \ref{fig:1}. \textit{Right axis:} The average vertical Mach number (green diamonds) as a function of initial inclination, calculated as the time average of $\mathcal{M}_{\rm z}$ over the whole evolution. The SMBH pair fails to merge at an inclination of 45$^{\circ}$ or more. Additionally, the average vertical Mach number very closely follows the trend of the pairing time with both increasing along with initial inclination. }
    \label{scatter}
\end{figure}

To quantitatively understand how sensitively $t_p$ depends on the initial inclination angle of the orbit, $t_p$ is plotted as a function of initial inclination for the same fiducial galaxy (Figure \ref{scatter}). The pairing time is approximately constant for small inclinations where the effects of vertical DF are very mild\footnote{The pairing time slightly decreases when the inclination is increased from 0 degrees to around 10 degrees due to numerical artifacts. There are steep gradients in the potential near the $z=0$ plane over the range $0 \textrm{ pc} <r <50$ pc due to the mass distribution of the bulge, which contains a singularity at the origin. The interpolating algorithm occasionally overestimates the gradient of the potential on coplanar orbits, leading to small injections of energy to the system, which slightly increases $t_p$ when $i_0 \lesssim 10$ degrees.}. However, at an inclination of about 20 degrees, the total pairing time undergoes a rapid increase until the pairing fails completely at an initial inclination of 45 degrees. SMBHs starting at inclinations beyond 45 degrees also fail to pair. Note that the decay times of the non-pairing orbits are potentially longer than 14 Gyr, but the computation caps $t_p$ to avoid possible infinite orbits. The reasoning for this extremely rapid change is the relationship between the strength of vertical DF and the Mach number which is explored in Figures \ref{25inc} and \ref{45inc}. When the initial average vertical Mach number is much greater than 1, it is very difficult to reduce vertical momentum since the strength of DF decays with increasing Mach number during supersonic motion.

\section{Analysis of Complete Parameter Suite}\label{Analysis}
To understand the impact of inclination in a wide variety of realistic post-merger galaxy models, the calculation is run for every combination of parameters listed in Table \ref{tab:params}, totaling 972 computations. The goal of this procedure is to identify any new relationships between different elements of the galaxy's structure and the initial inclination of the secondary. We identify any deviations from the results in Li20A which used many of the same parameters in its galaxy models but only for orbits coplanar with the galactic disk.

\begin{figure*}
    \centering
    \includegraphics[width=1.0\textwidth]{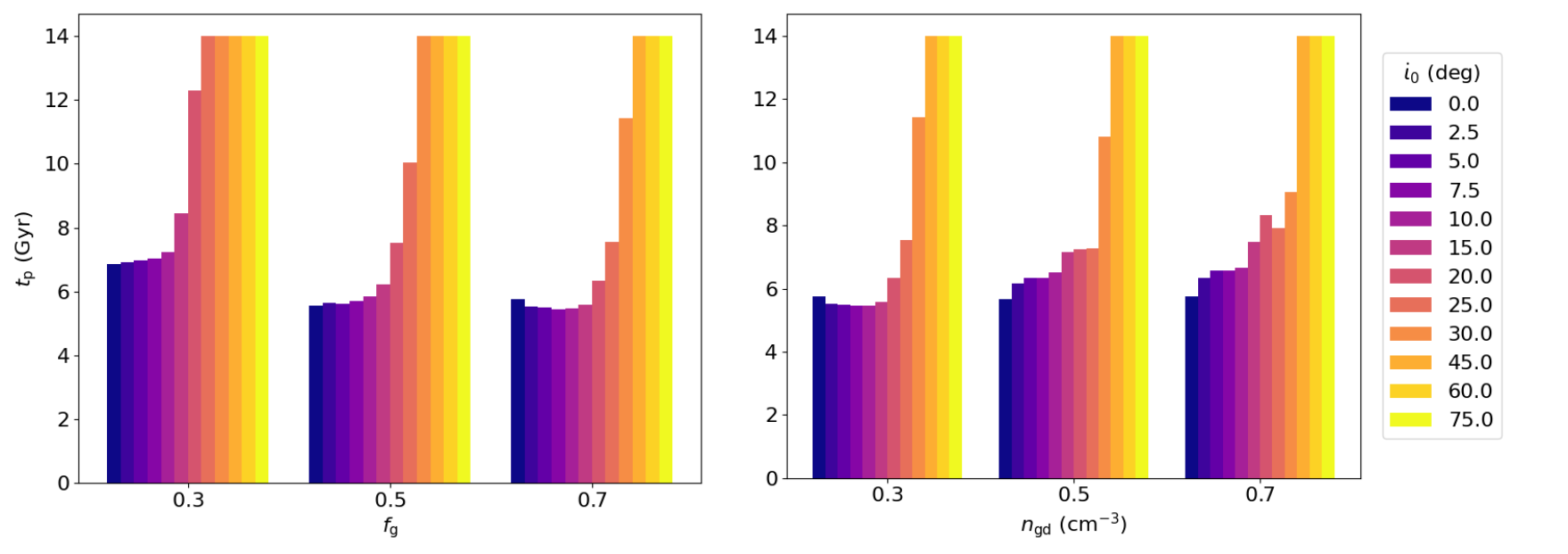}
    \caption{\textit{Left}: Pairing time of the SMBHs as a function of inclination angle for the three distinct gas fractions. Each inclination angle corresponds to a unique color. The parameters $n_{\rm gd}$, $v_{\rm g}$, and $M_1$ are fixed to the values as in the fiducial configuration from Figure \ref{fig:2}. The average pairing time generally decreases with increasing gas fraction. \textit{Right}: Pairing time as a function of inclination angle for three distinct central number densities $n_{\rm gd}$. All other parameters are fixed to the values in the fiducial configuration from Figure \ref{fig:2}. The relationship between central gas density and pairing time is somewhat weak, but pairing times generally increase with increasing $n_{\rm gd}$ due to the resulting increase in the secondary's Mach number, which weakens DF.}
    \label{ngd}
\end{figure*}

\begin{figure*}
    \centering
    \includegraphics[width=1.0\textwidth]{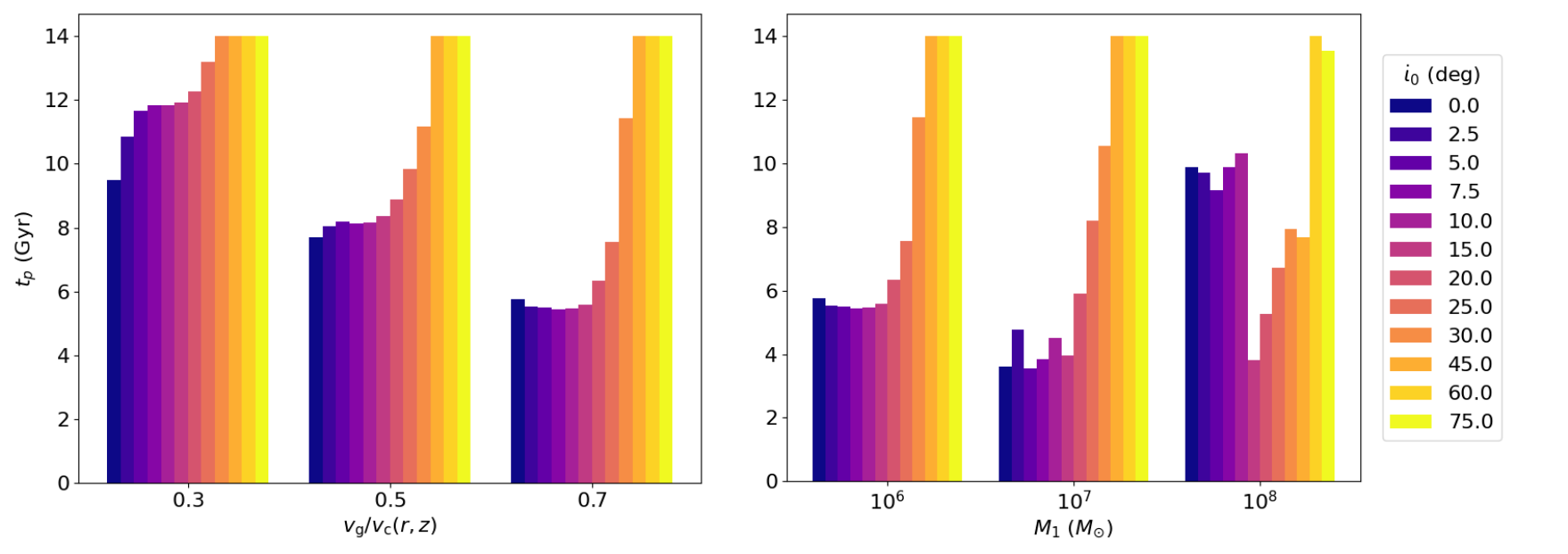}
    \caption{\textit{Left}: As in Figure \ref{ngd}, but with disk velocity $v_g$ as the parameter of interest. The pairing time diminishes with increased disk velocity. This is because the secondary is  moving at supersonic speeds, and DF is stronger for Mach numbers close to 1. \textit{Right}: As in Figure \ref{ngd}, but with primary mass $M_1$ as the parameter of interest. For some masses, $t_p$ can actually decrease with inclination contrary to intuition. This is because increasing the primary mass enhances the strength of DF but also increases the mass of the host galaxy due to the bulge mass varying directly with the primary mass.}
    \label{vg}
\end{figure*}

The impact of each parameter is addressed individually using Figures \ref{ngd}-\ref{vg}. These are multi-bar graphs that describe how $t_p$ jointly depends on inclination and the galaxy parameter of interest. The joint effect of changing gas fraction $f_{\rm g}$ and initial inclination $i_0$ is explored in the left panel of Figure \ref{ngd}. Three bar graphs corresponding to the the three values of $f_g$ describe how the pairing time evolves with increasing $i_0$ for each gas fraction. All other parameters are set equal to their values in the example from Figure \ref{fig:2}. For all three gas fractions, the same trend illustrated in Figure \ref{scatter} is observed. The time effectively remains constant for small initial inclinations but increases dramatically past a threshold inclination of 20 degrees, indicating vertical DF is no longer efficient. The key difference between each of the three distributions in the left panel of Figure \ref{ngd} is the magnitude of $t_p$. Generally, $t_p$ seems to decrease with increasing $f_{\rm g}$. This result is consistent with the findings in Li20A which determine that DF from gas disks is generally a more efficient drag force than DF from stellar disks when the Mach number is close to 1. Note that in all three cases, at very large inclinations, there is effectively zero probability of pairing, reflecting the previous observation that highly inclined orbits decay much slower due to higher vertical Mach numbers and decreased local stellar and gas densities for large $z$ (e.g. Figure \ref{45inc}). However, the critical angle at which pairing becomes impossible does change  with gas fraction. For example, at a gas fraction of $f_{\rm g} = 0.3$, pairings become impossible at initial inclinations larger than 20 degrees. However, at a gas fraction of $f_{\rm g} = 0.7$, pairings only become impossible at inclinations of 45 degrees or larger, so there is a noticeable disparity in the proportion of SMBH pairs that can decay at each gas fraction. 

The general relationship between pairing time and central gas density $n_{\rm gd}$ is similar to the trend for $f_{\rm g}$ and is illustrated in the right panel of Figure \ref{ngd}. Each of these bar graphs closely resembles Figure \ref{scatter}, and the pairing times generally increase with increasing $n_{\rm gd}$. While increasing central gas density would suggest that DF strength increase, the total mass increase of the galaxy resulting from this change also enhances the local circular speed, which can increase Mach numbers to the ranges where DF is inefficient.

The effect of varying disk rotation speed is analyzed in the left panel of Figure \ref{vg}. The differences in pairing times between each parameter value are much starker in this case. Each bar graph again exhibits a critical angle at which decay becomes impossible i.e. the bars flatten out at 14 Gyr. For $v_{\rm g} / v_c = 0.7$, we observe a trend similar to the bar graphs in Figure \ref{ngd} where $t_p$ remains approximately constant for small inclinations and then undergoes a dramatic shift. This is true for $v_{\rm g} / v_c = 0.5$ as well. The pairing times for smaller gas disk speeds tend to be longer because the secondary is generally moving at supersonic speeds. Therefore, if the gas disk is moving slowly, the Mach number of the secondary becomes too large for DF to efficiently drain its orbital energy. This explains why the pairing times increase substantially as the gas disk velocity decreases. Even though the changes in $v_{\rm g}$ are of order unity, the changes in $t_p$ are very large since order unity changes in the Mach number have a significant effect on the strength of DF as evidenced by Equation \ref{gasdf}. 

\begin{figure*}
    \centering
    \includegraphics[width=1.0\textwidth]{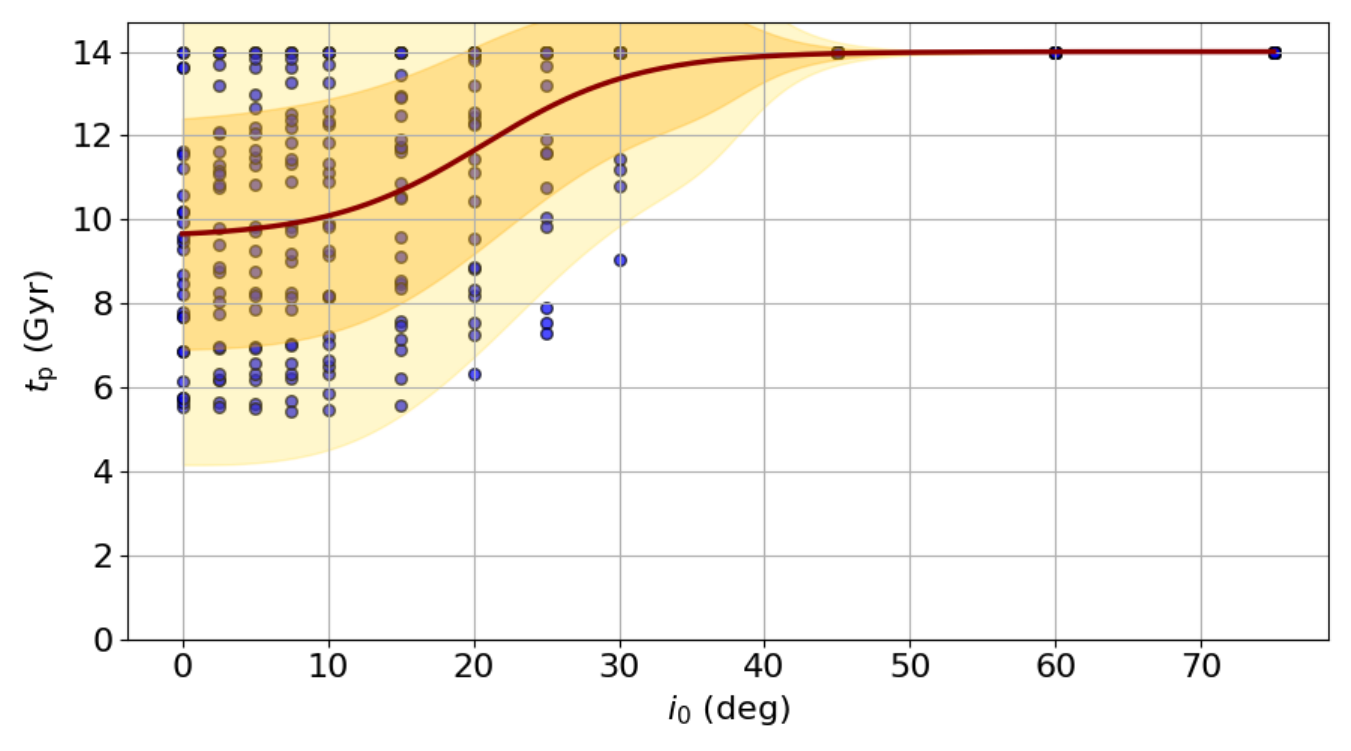}
    \caption{Distribution of all pairing times for a binary with primary mass of $10^6  M_{\rm \odot}$. The solid red line indicates a sigmoid function fit of the form in Equation \ref{sigma} to interpolate the average pairing time across all initial inclinations. The orange and yellow bands indicate $\pm 1\sigma$ and $\pm2 \sigma$ ranges of the data respectively. The bands condense significantly for higher inclinations since higher inclinations tend to yield non-decaying binaries (i.e., pairing times longer than 14 Gyr). Exact coefficients for the sigmoid fits are provided in Table \ref{paramtable}.}
    \label{m6}
\end{figure*}

\begin{figure*}
    \centering
    \includegraphics[width=1.0\textwidth]{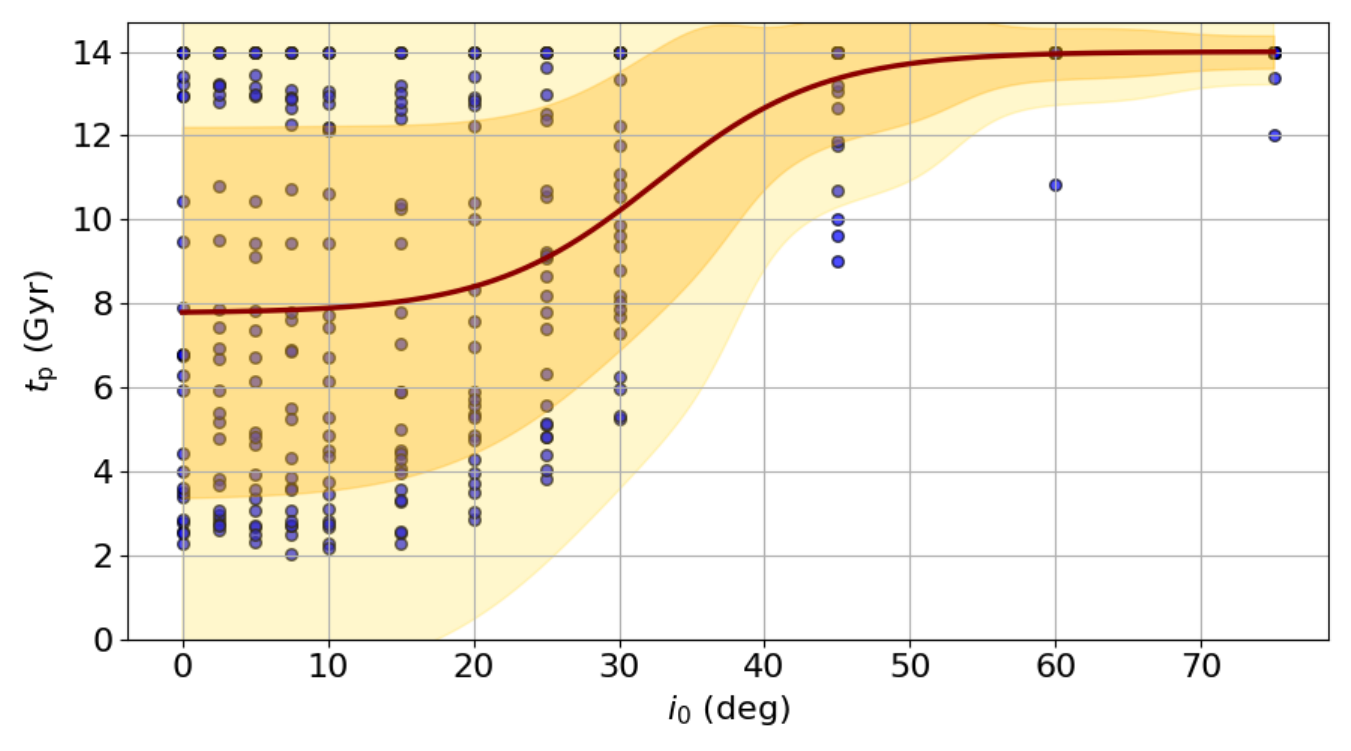}
    \caption{As in Figure \ref{m6}, but for a primary mass of $10^7  M_{\rm \odot}$. The relation between the pairing time and initial inclination is similar to that depicted in Figure \ref{m6} except pairing times are shorter on average. This is to be expected since the strength of DF varies with the square of the mass of the secondary.}
    \label{m7}
\end{figure*}

\begin{figure*}
    \centering
    \includegraphics[width=1.0\textwidth]{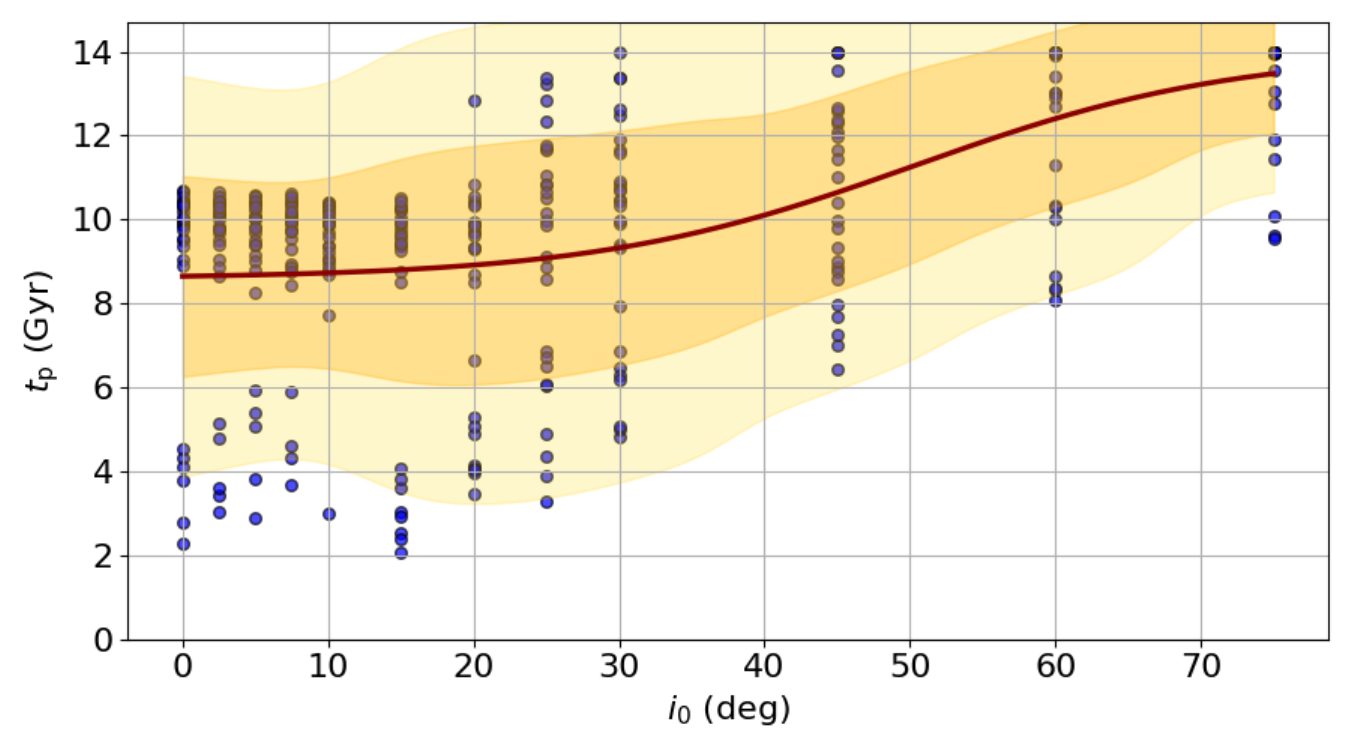}
    \caption{As in Figure \ref{m6}, but for a primary mass of $10^8  M_{\rm \odot}$. For $i_0 \lesssim 20^{\circ}$, there is a gap in the data points between short and long pairing times ranging from about 6-8 Gyr -- a distinct feature of orbits with a very massive primary. The data points with very short pairing times below this gap also correspond to galaxies with $v_{\rm g} = 0.7v_c$. As indicated in Figure \ref{vg}, the pairing time is very sensitive to disk velocity since a higher $v_{\rm g}$ generally decreases the Mach number of the secondary to the regime where gas DF is more efficient. This is especially important since orbits with higher mass primaries are generally dominated by DF in the stellar bulge, which is roughly constant across different inclinations due to the more spherical nature of the bulge.}
    \label{m8}
\end{figure*}

\begin{figure*}
    \centering
    \includegraphics[width=1.0\textwidth]{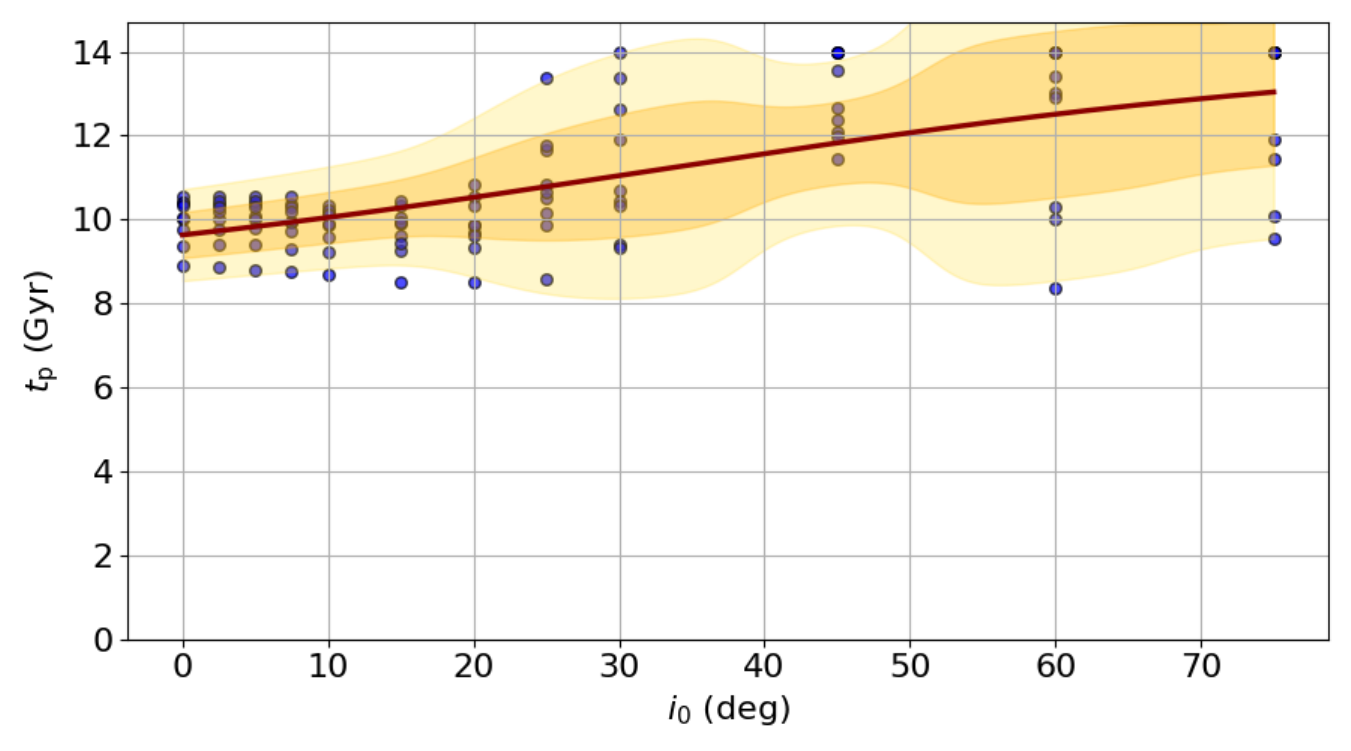}
    \caption{As in Figure \ref{m8}, but with the data restricted to include galaxies with only $v_g /v_c = 0.3$. As expected of slower rotating galaxies, $t_p$ is consistently high due to the inefficiency of DF at high Mach numbers. Unlike the full distribution of pairing times in Figure \ref{m8}, there are no pairing times lower than 8 Gyr among the slow rotating galaxies. The spread of pairing times is also substantially tighter as a result of this constraint.}
    \label{m83}
\end{figure*}

\begin{figure*}
    \centering
    \includegraphics[width=1.0\textwidth]{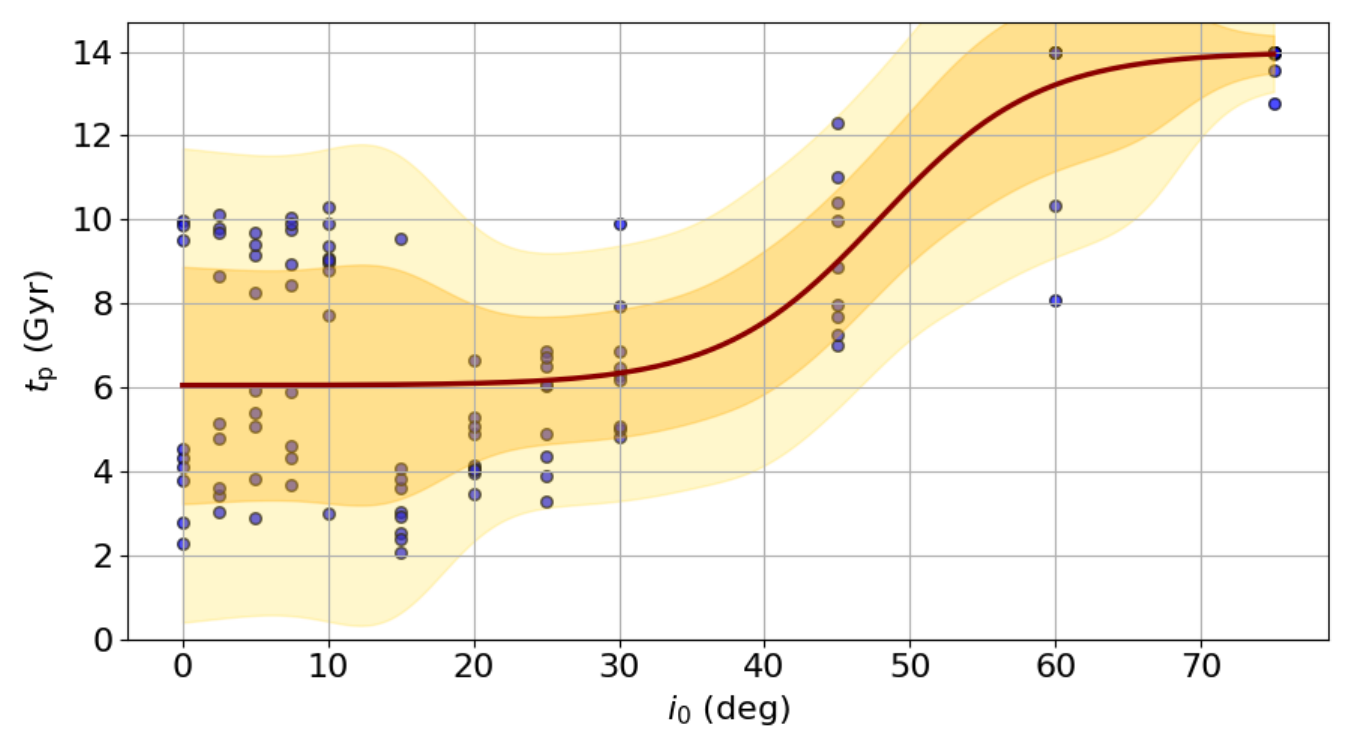}
    \caption{As in Figure \ref{m8}, but with the data restricted to include galaxies with only $v_g /v_c = 0.7$. The average pairing times in this case are significantly shorter than the previous distributions due to the combination of high disk speed and high SMBH mass, maximizing the efficiency of DF. }
    \label{m87}
\end{figure*}
The right panel of Figure \ref{vg} signifies that the primary mass is the most sensitive parameter of the suite, causing substantial changes in the pairing time when manipulated. There are several reasons for the pairing time to have such a strong relationship with binary mass. All forms of DF are proportional to the square of the mass of the secondary, so pairing times generally decrease with increasing mass. The mass of the primary also dictates the mass of the bulge, and therefore, the magnitude of stellar bulge DF. For binary masses of $10^6 M_{\rm \odot}$ and $10^7 M_{\rm \odot}$, the pairing time evolves with inclination in a manner consistent with Figure \ref{scatter}. Increasing the mass from $10^6$ to $10^7 M_{\rm \odot}$ also results in a general decrease of the pairing time as expected. 

With a binary mass of $10^8 M_{\rm \odot}$, however, the shape of the bar graph shifts considerably. Elevating the mass of the binary actually \textit{increases} the pairing time at low inclinations, and increasing the initial inclination from 15 degrees to 20 degrees \textit{decreases} the pairing time. This is a counterintuitive result, but it can be understood as follows. Increasing the mass of the central primary results in an increase in the mass of the central bulge, $M_b$. In a bulge dominated scenario, the circular velocity of the secondary is on the order of $\sqrt{GM_b / r}$, which means that increasing the mass of the bulge by 100 times increases the circular speed by a factor of about 10 in a bulge dominated scenario. This markedly increases the Mach number. Even though the DF strength is scaled up by the increased mass of the binary, the increased Mach number counteracts this effect and significantly diminishes the strength of DF at low inclinations. The abrupt decrease in pairing times noticeable at the inclination of 20 degrees arises because this effect is offset by two other competing ones. With increasing height from the plane of the disk, the secondary samples a shallower gravitational potential and its speed decreases to the point where DF is efficient again. In addition, inclining the orbit decreases the velocity component in the plane of the galaxy and introduces a velocity component in the perpendicular direction, boosting the efficiency of DF in both directions. 

Figures \ref{ngd} and \ref{vg} demonstrate the relative importance of each of the parameters in predicting pairing times. The pairing time is weakly affected by changes in gas fraction and central gas density. The mass of the SMBH primary and disk rotation speed have the greatest impacts on pairing time. The disparity in pairing times between the fastest rotating disks and the slowest rotating disks can reach several Gyr. The relationship between pairing time and primary mass is more complicated, but more massive SMBHs frequently take less time to pair than less massive SMBHs.

\section{Discussion}
\label{sec:discuss}
\subsection{Estimating SMBH Pairing Times}

The aim of this study is to quantify the degree to which orbital inclination influences pairing times when compared to orbits in the galactic plane. The primary result of the analysis is that larger inclinations generally increase the pairing time of an orbit due to the secondary traveling through less dense areas of gas and stars, decreasing the strength of DF. In cases of low to moderate inclination, vertical DF will cause the inclination to decay to 0 and regress the system to an uninclined orbit (e.g. Figure \ref{25inc}). For higher inclinations, vertical DF is extremely inefficient and becomes incapable of dragging the secondary to the galactic plane (e.g. Figure \ref{45inc}), resulting in many orbits failing to pair within a Hubble time. The data from the calculations may be used to support predictions for the populations of dual-AGNs in a variety of observed galactic configurations and may serve to motivate future observational surveys of precursors to LISA and PTAs sources. Though these calculations do not explicitly evolve the SMBH pair to the GW emitting regime, since DF is predicted to be the determining factor of the total decay time of an SMBH pair from merger to coalescence \citep[][]{2022ApJ...933..104L}, these results are important for understanding the expected rates of GW events associated with SMBH binaries.

\begin{table}[t]
    \centering
    \renewcommand{\arraystretch}{1.5}
    \begin{tabularx}{0.47\textwidth}{>{\centering\arraybackslash}X c c c}
        \toprule
        \textbf{$M_1$} & $\alpha$ & $\beta$ & $\gamma$ \\
        \midrule
        $10^6\, M_\odot$                       & $0.19$ & $20.69$ & $4.43$ \\
        $10^7\, M_\odot$                       & $0.17$ & $32.55$ & $6.23$ \\
        $10^8\, M_\odot$                       & $0.09$ & $50.41$ & $5.41$ \\
        $10^8\, M_\odot$ $(v_g/v_c = 0.3)$     & $0.04$ & $31.88$ & $5.71$ \\
        $10^8\, M_\odot$ $(v_g/v_c = 0.7)$     & $0.18$ & $47.93$ & $7.95$ \\
        \bottomrule
    \end{tabularx}
    \caption{Fitting parameters used to construct Figures \ref{m6}, \ref{m7}, \ref{m8}, \ref{m83}, and \ref{m87}. Each regression is a sigmoid fit of the form given in Equation \ref{sigma}.}
    \label{paramtable}
\end{table}

Figures \ref{m6}-\ref{m8} show fits to the pairing time as a function of $i_0$ for the three values of $M_1$. The fitting function is given by Equation \ref{sigma},
\begin{equation}\label{sigma}
    t_p(i_0) = \frac{\gamma}{1+e^{-\alpha(i_0-\beta)}} + 14-\gamma
\end{equation}
where the parameters $\alpha$, $\beta$, and $\gamma$ are found in Table \ref{paramtable}, and is demonstrated by the solid red line in all 3 figures. For simplicity, the fits are performed with all galaxy parameter values weighted equally. The distribution of parameters is not based on cosmological distributions of galaxy properties, so a Gaussian kernel is applied to the data to account for the spread in pairing times caused by the range of parameters. The orange and yellow bands represent the $\pm 1\sigma$ and $\pm2\sigma$ bands respectively. 

When $M_1 = 10^6 M_{\odot}$ (Figure \ref{m6}), the variation in pairing times is very large for small inclinations and ranges between approximately 6 and 14 Gyr. The bands condense as the inclination increases since higher inclination almost always predict pairing times greater than a Hubble time. The computation caps the pairing time at 14 Gyr, so all data points end up converging on this value. An important result of this plot is the existence of a critical non-pairing inclination for all pairs with $10^6 M_{\rm \odot}$ primaries. Therefore, an implication of the model is that pairing within a Hubble time is impossible for mergers that occur with initial inclinations greater than roughly 40 degrees for galaxies with a central SMBH on the order of $10^6 M_{\rm \odot}$.

Figure \ref{m7} is an identically constructed distribution plot for galaxies with $M_1 = 10^7 M_{\rm \odot}$ and reflects the same behavior as Figure \ref{m6}, except the bands are wider and minimum pairing times tend be shorter on average. This aligns with expectations due to the increased efficiency of DF for larger masses. Unlike the $10^6 M_{\rm \odot}$ case, Figure \ref{m7} exhibits pairing at all inclinations, though it still follows the same general trend of decreasing probability for higher $i_0$.

In Figure \ref{m8}, the predicted pairing times are analyzed for galaxies with a central SMBH of $10^8 M_{\rm \odot}$. There are several notable disparities between this plot and the two previous ones. For low inclinations, effectively all of the $10^8 M_{\rm \odot}$ computations result in pairing, which aligns with the prediction from Li20A that higher mass binaries generally have a higher probability of pairing. However, the variation in pairing times is significantly reduced at low inclinations for the $10^8 M_{\rm \odot}$ mass primary with most pairing times clustered around 10 Gyr. This seems to be a consequence of the bulge mass scaling with the mass of the central primary. The bulge is almost spherically symmetric, so increasing the mass of the bulge generally reduces the impacts of inclination on the system since the galaxy becomes more spherical. For low inclinations, there is also a substantial bimodality in the pairing times with a large gap between clusters of data. This gap occurs due to the variation in disk speed with shorter pairing times corresponding to faster rotating disks. As the inclination increases and the relative importance of the gas disk diminishes, this gap closes. Finally, note that it is always possible to observe pairing at any inclination when $M_1 =10^8 M_{\rm \odot}$ unlike the $10^6 M_{\rm \odot}$ mass case. 

To highlight the extreme degree to which disk velocity affects the pairing time in the $10^8 M_{\odot}$ case, we separately plot the distributions of $10^8 M_{\odot}$ pairing times for galaxies with only $v_g / v_c = 0.3$ or $v_g / v_c = 0.7$ in Figures \ref{m83} and \ref{m87} respectively. In Figure \ref{m83}, we note that $t_p$ is always greater than 8 Gyr and that the error bands are much tighter than in previous distributions. This is due to the inefficiency of DF at very low disk rotation speeds where the Mach number is highest. Figure \ref{m87} exhibits the distribution with the shortest pairing times on average due to the high SMBH mass and the high disk velocity. These conditions are the most conducive to maximizing DF efficiency. Figure \ref{m83} and \ref{m87} also confirm that the large gap in pairing times from Figure \ref{m8} is indeed due to variations in the disk speed.

\subsection{Impact of Simplifying Assumptions}\label{simplifying}

The semi-analytic approach is well-adapted to analyze a large number of SMBH pairs with minimal computational burden compared to more sophisticated numerical schemes. However, a necessary tradeoff is the imposition of extra assumptions that may diminish the accuracy or versatility of some of the calculations. We address the most important assumptions in this section.

One of the novelties in the approach to this problem is the use of a numerical Poisson solver to compute the potential of the post-merger galaxy, which would otherwise be too complicated to express in an analytic form. However, the numerical scheme only works under the assumption of axisymmetry, which restricts the types of structures that can be included in these calculations. In particular, the model cannot confidently predict the behavior of SMBH pairs in the presence of bars or spirals which break the symmetry of the setup. A prospective solution to this would be to upgrade the Poisson solver to a 3-dimensional box of dimensions $N \times M \times P$. This would naturally increase the computation time significantly. However, there are a number of ways to upgrade the Poisson solver to deal with the extra computational load such as by using a multigrid scheme or through parallelization \citep[e.g.][]{2011JCoPh.230.4756G}.

There are a few assumptions made in the calculation of the dynamics that may alter the computed pairing times. We assume a ``bare" secondary with no nuclear cluster of stars and gas. \cite{2017MNRAS.464.3131K} demonstrate that there is a high probability the secondary loses most of its stellar cluster by the time it reaches a separation of 1 kpc, so this is a plausible assumption. The pairing time would decrease if the secondary maintained a stellar cluster of comparable mass since DF would become more efficient. The accretion rate of the secondary is also ignored during its orbit which would cause it to increase in mass as it decayed. In principle, an increase in mass should lead to stronger DF and shorter inspiral times. However, when introducing accretion effects, the binary dynamics can be affected by the emitted radiation. For example, while increasing mass may strengthen DF, the simultaneous introduction of radiative feedback may counteract this effect and accelerate the SMBH \citep{2017ApJ...838..103P}.  We do not model the effect of radiative feedback here and instead focus on the impact of orbital inclination.

A crucial mechanism in understanding how inclination affects the inspiral time was vertical DF. For vertical DF from stars, we were able to adapt the models by \cite{AM2012}, but there exists no definitive treatment of vertical gas DF to the best of the authors' knowledge. A key assumption was the adaptation of the model by \cite{O1999} to approximate the drag in the vertical direction. However, the dynamical calculations strain some of the details of this model, particularly the assumption that the secondary moves through a uniform field of gas and does not change direction during its orbit. The implementation of this model to approximate vertical DF from gas is a suitable order of magnitude calculation, but the model's accuracy is not sufficient near the vertical turning points of the orbit where the $z$-velocity of the secondary is close to 0, breaking down key assumptions of most DF analyses.

There are several dynamical scenarios that were not tested through the parameter suite. In particular, two key parameters that are not addressed but which were present in the analysis of Li20A are the initial eccentricity of the orbit and the rotation direction of the gas disk. Though these parameters are not probed explicitly, the analyses of the previous sections offers theoretical explanations for the dynamics in these cases. For very eccentric orbits, shorter pairing times are expected compared to less eccentric orbits in the same conditions since the small pericenter of these types of orbits leads to increased exposure of higher densities of gas and stars, increasing the strength of DF. For retrograde orbits, longer pairing times are generally expected due to the increased Mach number of the secondary. In most configurations investigated here, the secondary travels at supersonic speeds, even in prograde orbits. Therefore, the secondary will move even faster relative to the galactic disk on a retrograde orbit, which drastically decreases the efficiency of DF.

Finally, note that dark matter halos, a key structure in the understanding of galactic dynamics, are omitted in the analysis. The primary reason for this omission is the size of the computing space which is only a few kpc in length. Variations in the potential of the halo are very small compared to variations in the potential of the bulge and disks on this scale \citep{BT1987}. This is evident when analyzing the scale radii of these structures. For example, the scale radius of an NFW halo profile is on the order of $10$ kpc \citep[][]{1997ApJ...490..493N, Klypin2002}, which is beyond the scope of the simulation.

\section{Conclusions}
\label{sec:concl}
We present the results of a 3D dynamical model for calculating the pairing times of SMBH pairs in post-merger remnant galaxies. The model provides a novel way of analyzing SMBH pairs under the influence of stellar and gas DF during inclined orbits within a few kpc of the center of the galactic disk. The model is tested on a suite of 972 distinct simulations to allow us to predict the pairing times of SMBH pairs in a wide variety of galactic environments. The key takeaways from the analysis are listed:

\begin{enumerate}
    \item Pairing times are sensitive to the inclination of an SMBH pair orbit. Pairing times generally \textit{increase} with increasing initial inclination due to the dependence of DF on the local density of gas and stars, an effect demonstrated in Figure \ref{scatter}. DF is the primary driver of SMBH pairing, and the strength of DF diminishes in areas of low gas and star density. Therefore, for flattened systems like disk galaxies, inclination drives the secondary away from areas of high mass density causing the overall average effect of DF to weaken.

    \item For many galaxies, there is a critical angle of initial inclination for which pairing is impossible in the span of a Hubble time. Inclined orbits often decrease in inclination due to the effect of vertical DF, which gradually drags the secondary back into the plane of the disk (e.g., Figure \ref{25inc}). This suggests that dAGN in gas-rich merger remnants may preferentially appear in coplanar configurations, offering a valuable observational constraint for future surveys. However, at these critical initial inclinations, the vertical speeds of the secondary are much higher than the local sound speed which increases the Mach number of the secondary well beyond the regime where vertical DF is efficient. The secondary becomes stuck at large inclinations in these cases (e.g., Figure \ref{45inc}). Therefore, inclination generally reduces the probability of pairing. 

    \item The introduction of inclination also affects how pairing times depend on other aspects of the galaxy's structure including its disk speed $v_{\rm g}$, its central gas density $n_{\rm gd}$, its gas fraction $f_{\rm g}$, and the mass of the central SMBH $M_1$. Variations in $f_{\rm g}$ and $n_{\rm gd}$ generally result in minor shifts in the pairing time when compared to the effects of altering the initial inclination of the orbit (e.g., Figure \ref{ngd}). The pairing time depends more sensitively on the disk speed $v_{\rm g}$ and the central mass $M_1$ (e.g., Figure \ref{vg}). The pairing time is minimized when the gas disk is spinning at its fastest ($v_{\rm g} = 0.7v_c$ in this case).
    
    \item Pairing times vary dramatically with changes in $M_1$ since the mass of the central bulge is dictated by $M_1$. Pairing times are generally shorter for galaxies with $M_1 = 10^7 M_{\rm \odot}$ compared to galaxies with $M_1 = 10^6 M_{\rm \odot}$ (see Figures \ref{m6} and \ref{m7}). However, for galaxies with central SMBH mass $10^8 M_{\rm \odot}$, the pairing times increase for small initial inclinations and then drop dramatically at intermediate inclinations (e.g., Figure \ref{m8}). This strange behavior is due to the dramatic change in structure of the galaxy which becomes bulge dominated for $M_1 = 10^8 M_{\rm \odot}$. In this regime, Mach numbers are too high at small inclinations, so DF is generally not as strong as it would be in a less massive bulge scenario. Ultimately, the shortest inspiral times generally occur for SMBH pairs at small inclinations in galaxies with fast rotating disks.
\end{enumerate}

Future extensions of this work could include incorporating additional physical processes such as gas accretion onto the SMBHs or feedback from AGN activity, both of which may further modify pairing times. Expanding the model to larger samples of cosmologically motivated galaxy merger remnants would also help bridge the gap between simulations and realistic merger histories. More broadly, the framework developed here may provide useful input for studies of dual AGN demographics, the efficiency of SMBH coalescence in different environments, and predictions for gravitational wave event rates observable by missions such as LISA.
\\
\begin{acknowledgements}
The authors acknowledge support from NSF grant AST-2407658. DRB is also supported by NASA award 80NSSC24K0212.
\end{acknowledgements}

\bibliographystyle{aasjournalv7}

\end{document}